\documentclass[aps,prd,reprint,showpacs,superscriptaddress]{revtex4-1}
\usepackage{graphicx}
\usepackage{amssymb}
\usepackage{amsmath}
\usepackage{bm}
\usepackage{verbatim}
\usepackage{enumitem} 

\usepackage{pdfpages}

\newcommand{\im}{\Im}

\newcommand{\dert}{\partial_t}
\newcommand{\derx}{\partial_x}

\newcommand{\der}{\partial}
\newcommand{\dx}{\mathrm{d}x}

\newcommand{\dt}{\mathrm{d}t}

\newcommand{\ii}{\mathrm{i}}

\begin{document}

\title{Kinetics of Mobile Impurities and Correlation Functions in
  One-Dimensional Superfluids at Finite Temperature}

\author{M.~Arzamasovs}
\affiliation{School of Physics and Astronomy, University of Birmingham,
Edgbaston, Birmingham, B15 2TT, UK}
\author{F.~Bovo}
\affiliation{School of Physics and Astronomy, University of Birmingham,
Edgbaston, Birmingham, B15 2TT, UK}
\author{D.M.~Gangardt}
\affiliation{School of Physics and Astronomy, University of Birmingham,
Edgbaston, Birmingham, B15 2TT, UK}

\date{\today}

\begin{abstract}
We examine the hydrodynamic approach to dynamical correlations
in one-dimensional superfluids near integrability and calculate the
characteristic time scale, $\tau$, beyond which this approach is valid.
For time scales  shorter than $\tau$, hydrodynamics 
fails, and we develop an approach based on kinetics of fermionic quasiparticles 
described as mobile impurities. New universal results 
for the dynamical structure factor relevant to experiments in ultracold atomic gases
are obtained.

\end{abstract}


\maketitle


The understanding of many-body quantum dynamics is a central topic of
current research in ultracold atomic gases \cite{Dziarmaga2010,*Polkovnikov2011}.  
Of special interest are one-dimensional (1D) systems where 
interactions play a crucial role in their dynamics  as was demonstrated 
in recent experiments~\cite{Kinoshita06,AduSmith2013a}.  

These effects are accessible by measuring the dynamical correlations of
1D bosons, which in contrast to the static ones \cite{Cazalilla2011}, largely
remain an open theoretical problem \cite{Imambekov2012}.  One such correlation
function is the dynamical structure factor (DSF) defined as the Fourier transform
of the density-density correlation function (we use the units such that
$\hbar=1,\; k_B=1$ throughout the paper)
\begin{eqnarray}
  \label{eq:sqwdef}
  S(q,\omega) = \int\!\dx\,\dt\, e^{-i(qx-\omega t)} \langle \rho(x,t)\rho(0,0)\rangle\, .
\end{eqnarray}
This quantity is readily accessible by Bragg spectroscopy \cite{Ketterle1999},
and its measurement was recently reported for arrays of one-dimensional 
Bose gases \cite{Clement2011}.

As the calculation of the correlation functions, especially dynamical ones, is
still a formidable task even for integrable models, the problem is often
solved by employing a hydrodynamic description valid in the low-energy limit
\cite{HaldanePRL81,PopovBookFunctional}, which treats a system as a collection
of weakly interacting phononic modes. In the case of weakly interacting  
1D bosons at low enough temperatures $T<mc^2$,  where $c$
is the sound velocity and $m$ is the mass of particles, 
and for small  momenta $0<q<T/c$, the DSF is a narrowly peaked function 
around the line $\omega = cq$ 
\footnote{Here we concentrate on positive momenta $q>0$ and
  positive frequencies $\omega$. All the
  arguments are, of course, valid for the time-reversed situation $q\to -q$.
For positive $q$ at finite temperature, there is also an emission peak at 
$\omega=-cq$ exponentially suppressed for $T<mc^2$.},
suggesting that the main contribution into
$S(q,\omega)$ comes from phononic excitations.  The smearing of the phononic
peak was attributed to the phononic nonlinearity by Andreev \cite{Andreev1980} 
who used the hydrodynamic approach to obtain the typical width
\begin{eqnarray}
  \label{eq:dwAndreev}
  \delta \omega^{(B)}_q = 0.394 \left(1+\frac{n}{c}\frac{\der c}{\der n}\right) 
  \sqrt{\frac{T q^3}{m n}}\, ,
\end{eqnarray}
where $n$ is 1D density. 
The width in Eq.~(\ref{eq:dwAndreev}) 
has a 3/2 power law dependence on momentum, suggesting 
a relation to the Kardar-Parisi-Zhang (KPZ) 
model of interface growth \cite{Kardar1986,*Kriecherbauer2010,*Sasamoto2010a}, 
as  was elucidated and recently confirmed numerically by Kulkarni and Lamacraft~\cite{Kulkarni2013}. 

\begin{figure}
  \centering
 \includegraphics[width=\columnwidth]{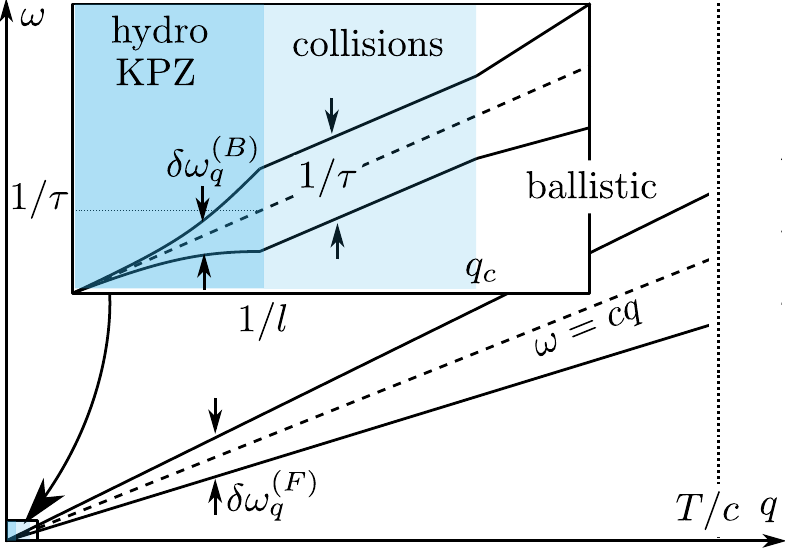}
   \caption{(Color online) Typical support in
     the $(q,\omega)$ plane of the dynamical structure factor $S(q,\omega)$ in
     different regimes. The dominant feature is the   
     ballistic behavior linear in momentum $q$ as in 
     Eq.~(\ref{eq:dwTG}). The hydrodynamic result Eq.~(\ref{eq:dwAndreev})
     shown in the inset 
     is only valid  for  momenta $q<1/c\tau$, where the
     collision time $\tau$ is given by Eq.~(\ref{eq:tauF}).  }
  \label{fig:sqw}
\end{figure}
Formally, Eq.~(\ref{eq:dwAndreev}) (which was
obtained under assumptions of the hydrodynamic description, i.e., in the
universal long-wavelength limit) is valid for all values of the interaction
\cite{Samokhin1998}, including the case of impenetrable 
Tonks-Girardeau (TG) limit. Conversely, in this regime, 
one can use the Bose-Fermi mapping
\cite{Girardeau60} representing excitations in terms of free fermions. This
leads to a different estimate for the width of DSF,
\begin{eqnarray}
  \label{eq:dwTG}
  \delta\omega^{(F)}_q \sim \frac{qT}{mc}\, . 
\end{eqnarray}
The physics behind Eq.~(\ref{eq:dwTG})
is that the density fluctuations are dominated by effective 
fermionic particle-hole
excitations rather than phonons. The fermions move
ballistically with velocity close to $c$ and the width (\ref{eq:dwTG}) is
just the uncertainty of the energy of a particle-hole pair
$\delta\omega_q^{(F)} =\delta c q$ resulting from the thermal uncertainty in
the Fermi velocity $\delta c \sim T/mc$ \cite{Imambekov2012}.  For small
momenta $q<T/c$, the energy uncertainty of a pair  exceeds considerably 
the one predicted by Andreev, 
$\delta\omega_q^{(F)}>\delta\omega_q^{(B)}$.

Away from the TG regime, the Bose-Fermi mapping can be generalized to any value of
interactions \cite{Cheon1999FermionBoson}. For this effective system of
interacting fermions, the linear momentum law, Eq.~(\ref{eq:dwTG}), still
holds with a proper redefinition of the Fermi velocity and mass
\footnote{Equation ~(\ref{eq:dwTG}) first appeared in M. Pustilnik, E. G. Mishchenko, L. I. Glazman, and A. V. Andreev, Phys. Rev. Lett. {\bfseries 91}, 126805 (2003)
  without derivation. The derivation identical to one reproduced here after 
Eq.~(\ref{eq:dwTG}) can be found in Ref.~\cite{Imambekov2012}. }.

We now examine the hydrodynamic approach of
Refs.~\cite{Andreev1980,Kulkarni2013} and argue that it can be applied only on
time scales longer than a certain microscopic time $\tau$ restricting the
hydrodynamic behavior of $S(q,\omega)$ to very small frequencies i.e., the
standard condition $\omega\tau <1$ \cite{lifshitz1981physical}.  The time
$\tau$ and corresponding length $l = c\tau$ can be identified with the lifetime and the mean free path, respectively, of fermionic excitations, which can
be defined away from the TG limit \cite{Rozhkov2005,*Rozhkov2008} and which
represent fast degrees of freedom. So, the apparent contradiction between the
results in Eqs.~(\ref{eq:dwAndreev}) and ~(\ref{eq:dwTG}) is naturally
resolved: for times shorter than $\tau$, the fermionic quasiparticles move
ballistically, and the spread in their energies is controlled by the thermal
uncertainty in their velocities leading to Eq.~(\ref{eq:dwTG}).  The DSF
for $\omega>1/\tau$ is then given by the standard fermionic
expression with renormalized mass and quasiparticle residue; see
Eq.~(\ref{eq:sqwFlargeq}) below. For long times, $t>\tau$, or small frequencies,
$\omega<1/\tau$, the fermionic quasiparticles thermalize, and the system enters
the hydrodynamic regime. The DSF acquires the form
given by Eq.~(\ref{eq:sqwkpz}) with width given by
Eq.~(\ref{eq:dwAndreev}). There is also an intermediate regime, where the
width of $S(q,\omega)$ is controlled by collisions between excitations; see
Eq.~(\ref{eq:sqwFsmallq}).  These regimes of DSF are
shown schematically in Fig.~\ref{fig:sqw}.

The exact microscopic calculations of $\tau$ can only be  envisaged for integrable
models, but the result $1/\tau=0$ is expected due to  infinitely many
integrals of motion preventing the system from thermalization. 
We consider a small perturbation of the Lieb-Liniger model and 
obtain the main result of this Letter, 
the expression for thermalization time scale,
\begin{eqnarray}
  \label{eq:tauF}
  \frac{1}{\tau} =  \frac{\pi^5}{128}\big[\Gamma'_0\big]^2\frac{
   T^7 }{{m^*}^2c^6}\, , 
\end{eqnarray}
where  the effective mass $m^*$ is 
given by Eq.~(\ref{eq:eff_mass}).  The time scale  $\tau$
depends strongly on the temperature $\tau\propto T^{-7}$. In addition, it is
proportional to the square of the parameter $\Gamma'_p=\der \Gamma_p/\der
p$. Here, $\Gamma_p$ is the amplitude of backscattering of phonons by a
fermionic quasiparticle with momentum $p$ introduced in
Refs.~\cite{Gangardt09,Gangardt2010Quantum,Schecter_Gangardt_Kamenev_2012}  
so that $\Gamma_p mc^2$ is the dimensionless small parameter of our
theory. It
depends on the fine details of the interactions between particles; in
particular, it vanishes identically for the Lieb-Liniger model
\footnote{A.S. Campbell, PhD Thesis, 2012; A.S. Campbell and D.M. Gangardt,
  unpublished}. For weakly interacting bosons with weak three-particle
interactions, we find $\Gamma'_0 =  -\alpha n/(48 m^2c^4)$, where 
$\alpha=\der(mc^2/n)/\der n$.  Below, we discuss the derivation of the above
results \footnote{See Supplemental Material at [url], which includes Refs. \cite{Wadkin-Snaith2012, Khodas2007, MatveevFurusaki2013, Kamenev, GradshteynRyzhik}, for details of mobile impurity theory, Keldysh technique and its application to the derivation of the Boltzmann equations for the system considered.}. \nocite{Wadkin-Snaith2012, Khodas2007, MatveevFurusaki2013, Kamenev, GradshteynRyzhik}

The hydrodynamic description of one-dimensional superfluids 
\cite{HaldanePRL81,PopovBookFunctional} is based on considering 
smooth configurations of the displacement 
$\vartheta$ and phase $\varphi$ fields related to deviations 
$\rho= \derx \vartheta/\pi$ of the density
from its thermodynamic average $n$ and superfluid velocity 
$u=\derx\varphi/m$. Using these variables, the low-temperature dynamics is then
governed by the Lagrangian density
\begin{eqnarray}
  \label{eq:lhydro}
  \begin{split}
  \mathcal{L}_\mathrm{hyd} &= -\rho\dot\varphi -
  \frac{n+\rho}{2m}\left(\derx\varphi\right)^2 \\
  &-\big[ e_0 (n+\rho) -
    e_0(n)-\mu\rho \big]\, ,
  \end{split}
\end{eqnarray}
where $e_0(n)$ is the ground state energy density and $\mu = \der e_0/\der n$
is the chemical potential.  
Expanding  Eq.~(\ref{eq:lhydro}) for small $\rho, u$, 
one obtains the quadratic phononic Lagrangian density
\begin{eqnarray}
  \label{eq:lll}
  \mathcal{L}_\mathrm{ph} =
  -\frac{1}{\pi}\derx\vartheta\dert\varphi -
    \frac{c}{2\pi K}\left(\derx\vartheta\right)^2
    - \frac{cK}{2\pi}\left(\derx\varphi\right)^2\, 
\end{eqnarray}
depending on the Luttinger parameter $K=\pi n/mc$.  For
bosons with short-range interactions, $K\ge 1$. In the
weakly interacting regime, $K\to\infty$, while $K=1$ corresponds to the
TG gas of hard-core bosons. 
It is customary to define the right and left chiral fields 
\begin{eqnarray}
  \label{eq:chi}
 \chi_\pm = \vartheta/\sqrt{K} \pm \varphi \sqrt{K} 
\end{eqnarray}
in terms of which the phononic Lagrangian separates 
\begin{eqnarray}
  \label{eq:lchiral}
      \mathcal{L}_\mathrm{ph} =\sum_{\nu=\pm}
   \frac{1}{4\pi} \chi_\nu^\dagger\left(\nu 
  \dert\derx+c\derx^2\right)\chi_\nu\, . 
\end{eqnarray}

The next (cubic) order the expansion of Eq.~(\ref{eq:lhydro}) contains a 
nonlinear coupling between phonons,
\begin{eqnarray}
  \label{eq:lnonlin}
  \begin{split}
  \mathcal{L}_\mathrm{ph}' &= \frac{\alpha}{6}\rho^3+
  \frac{1}{2m}\rho\left(\derx\phi\right)^2 \\
  &=\frac{1}{12\pi m^*}  
  \left[\left(\derx\chi_+\right)^3 + \left(\derx\chi_-\right)^3\right]
  +\ldots\, .
  \end{split}
\end{eqnarray}
The omitted terms  describe the interactions between phonons of different
chirality. They  can be safely neglected for low energies, as
the interaction time between phonons moving in opposite directions
is small. In Eq.~(\ref{eq:lnonlin}), we have introduced the effective mass 
\begin{eqnarray}
  \label{eq:eff_mass}
 \frac{1}{m^*} = \frac{1}{2m\sqrt{K}}\left(1+\frac{n}{c}\frac{\der c}{\der n}\right)
\end{eqnarray}
(c.f. Ref.~\cite{Pereira2006}).  
In the TG regime, $m/m^*= 1$, 
whereas in the weakly interacting regime, 
$m/m^* = 3/(4\sqrt{K})$ \cite{Imambekov2012}.

The total Lagrangian density
$\mathcal{L}_\mathrm{ph}+\mathcal{L}'_\mathrm{ph}$ generates equations of motion
for the right-moving fields
\begin{eqnarray}
  \label{eq:kpz}
  \left(\dert+c\derx\right)\chi_+ = \frac{1}{2m^*}\left(\derx\chi_+\right)^2 +
  D\derx^2\chi_+ + \xi\, ,
\end{eqnarray}
and similarly for the left-moving ones. In Eq.~(\ref{eq:kpz}), 
we have also added terms describing dissipation and
thermal white noise with zero mean and 
$\overline{\xi(x,t)\xi(x',t')} = (4\pi DT/c)
\delta(x-x')\delta(t-t')$ originating from coupling to a yet unspecified thermal
bath with temperature $T$.  Equation ~(\ref{eq:kpz}) can be identified with the
celebrated KPZ equation for the dynamics of interface growth
\cite{Kardar1986,*Kriecherbauer2010,*Sasamoto2010a}.  Its DSF was obtained in terms of the universal function $\mathring{f}(s)$
describing the scaling limit of the polynuclear growth model 
\cite{Prahofer2004}.  The function
$\mathring{f}(y)$ can only be determined numerically and has a peaked form
with height and width of order unity.  
Expressing the density in terms of the chiral 
field using Eq.~(\ref{eq:chi}), the result can be written as
\begin{eqnarray}
  \label{eq:sqwkpz}
  {S}(q,\omega)  = 
\frac{T}{c} \frac{K}{2\pi^2} \frac{1}{\delta\tilde\omega^{(B)}_q} \,
  \mathring{f}\left(\frac{\omega-cq}{\delta\tilde\omega^{(B)}_q}\right)\, .
\end{eqnarray}
Its typical width is given by  
\begin{eqnarray}
  \label{eq:domega_KPZ}
\delta\tilde\omega^{(B)}_q = \sqrt{\frac{2Tq^3}{{m^*}^2 c}}  
\end{eqnarray}
in full agreement with Andreev's results \footnote{Although the
  expressions ~(\ref{eq:dwAndreev}) and ~(\ref{eq:domega_KPZ}) seem
  different, using the effective mass (\ref{eq:eff_mass}), one obtains
  $\delta\tilde\omega^{(B)}_q/\delta\omega^{(B)}_q = (2\pi)^{-1/2}/0.394\simeq
  1.0125$}.  The numerical calculations in Ref.~\cite{Kulkarni2013} based on the
Gross-Pitaevskii equation confirm
results~(\ref{eq:sqwkpz})~and~(\ref{eq:domega_KPZ}).

While the calculations leading to Eqs.~(\ref{eq:sqwkpz}) and ~(\ref{eq:domega_KPZ})
are purely classical, based on the hydrodynamic Lagrangian
(\ref{eq:lhydro}), the calculations of Ref.~\cite{Rozhkov2005,*Rozhkov2008} 
leading to Eq.~(\ref{eq:dwTG})  are  
essentially quantum, based on the quantization of bosonic fields using
the commutation relations 
$  [\hat{\phi}(x),\hat{\rho}(x')] = \ii\delta(x-x')$ 
and the introduction of the fermionic operator
\begin{eqnarray}
  \label{eq:fermi_ops}
  \hat{\psi}_{+} (x) \sim e^{-\ii \hat{\chi}_+(x)} 
  = e^{-\ii \sqrt{K}\hat{\varphi}(x) - \ii \hat{\vartheta}(x)/\sqrt{K}}\, ,
\end{eqnarray}
and similarly for $\hat\psi_-$ \cite{Imambekov2012}. The density can be
expressed in terms of these operators as $\hat\rho =
\sqrt{K}(\hat\psi_+^\dagger\psi^{\phantom{\dagger}}_++ \hat\psi_-^\dagger\psi^{\phantom{\dagger}}_-)$, though only
right-moving fermions contribute in our case.  
The fermionic operators (\ref{eq:fermi_ops}) create ``kinks'' of magnitude $\sqrt{K}$ and $1/\sqrt{K}$ in
the fields $\vartheta$ and $\varphi$, correspondingly. Such kink configurations can be described semiclassically as
``mobile impurities'' (even in the absence of foreign particles) and were
recently studied in Ref.~\cite{Schecter_Gangardt_Kamenev_2012}. 

In the integrable case of the Lieb-Liniger model, one can associate the
impuritylike excitations with two excitation modes, Lieb~I, $E_I(p)$ and 
Lieb~II, $E_{II} (p)$   predicted  by the Bethe
ansatz solution \cite{Lieb_Liniger_1963,*Lieb_1963}. 
For small momenta $E_{I,II}(p) = cp \pm p^2/2m^*$
\cite{Imambekov2012} allowing the interpretation of $E_{I}$ as a
particle and $E_{II}$ as a hole, the excitations created on top of the Fermi 
ground state filled for $p<0$ using the standard prescription 
$E_{I,II} (p) = \pm \varepsilon_+ (\pm p)$. The dispersion law of the 
right-moving fermions \footnote{For left-moving fermions we have
  $\varepsilon_-(p) = -cp +p^2/2m^*$}, 
\begin{eqnarray}
  \label{eq:dispersion}
  \varepsilon_+(p)= cp +\frac{p^2}{2m^*}\, 
\end{eqnarray}
contains the parameters $c$ and $m^*$, the same as the ones entering the hydrodynamic
Lagrangian (\ref{eq:lchiral}) and ~(\ref{eq:lnonlin}), as both phononic and
fermionic descriptions
must provide the same equilibrium properties of the system
\cite{Pereira2006,Imambekov2012}.

In the TG regime, the fermions are exact excitations, and Eq.~(\ref{eq:dispersion})
is valid for all momenta. It remains valid even in the weakly interacting
regime but for restricted momenta $|p|<p_\mathrm{G} = (m/m^*) mc
\sim mc/\sqrt{K}$ \cite{Imambekov2012}. This puts a further limitation on the
temperatures $T<T_\mathrm{G} = cp_\mathrm{G} \sim mc^2/\sqrt{K}$ for the validity
of our approach. For $|p|>p_\mathrm{G}$ the excitations can be treated
semiclassically: the Lieb I and II modes cross over into the Bogoliubov mode and
gray soliton mode respectively \cite{KulishManakovFaddeev76}.

In the integrable case, the fermionic excitations are exact eigenstates of the
system, so they have an infinite lifetime.  Away from integrability, they
experience backscattering from phononic excitations providing the mechanism for 
their decay. In the low-temperature limit, the backscattered phonons have very
small momenta to transfer to the impurity, and the corresponding amplitude
reduces to a function $\Gamma_p$ of the impurity momentum $p$ only and can be
calculated phenomenologically from the dependence of the dispersion 
(\ref{eq:dispersion})  on the background density 
$n$ \cite{Schecter_Gangardt_Kamenev_2012}.  
If, in addition, the momentum of the
impurity is small, the amplitude $\Gamma_p \simeq \Gamma'_0 p$, so the
backscattering is characterized by only one parameter, $\Gamma'_0$.

To derive the DSF,  
we assume the system to be only slightly
nonintegrable, so we can rely on perturbation theory in $\Gamma'_0$ 
and employ the picture of the near integrable Bose liquid as a collection 
of phonons described by Eq. ~(\ref{eq:lchiral}) 
and fermionic quasiparticles treated as a dilute gas of 
mobile impurities with a Lagrangian density 
\begin{eqnarray}
\label{eq:lagrimp2nd}
  \mathcal{L}_\mathrm{F}^{+} = 
  \psi_+^\dagger\bigg[\ii\dert-\varepsilon_+(-\ii\derx)\bigg]\psi_+ \, . 
\end{eqnarray}
The term describing low-energy interactions between fermions and phonons can be
written as
\begin{eqnarray}
  \label{eq:lint}
  \mathcal{L}^+_\mathrm{ph-F} = -\frac{c\Gamma'_0}{2\pi m^*}\derx\chi_+\derx\chi_-
  \left(\psi^\dagger_+\overleftrightarrow{\derx^2}\psi_+\right)\, ,
 \end{eqnarray}
where we have introduced the symmetrized operator
$\overleftrightarrow{\derx^2} =(1/4)(
\overleftarrow{\derx^2}-2\overleftarrow{\derx}\overrightarrow{\derx}+\overrightarrow{\derx^2})$
with arrows denoting the derivatives acting on the  operators to the
left or right. Operator hats are omitted  for clarity. 

\begin{figure}
  \centering
 \includegraphics[width=\columnwidth]{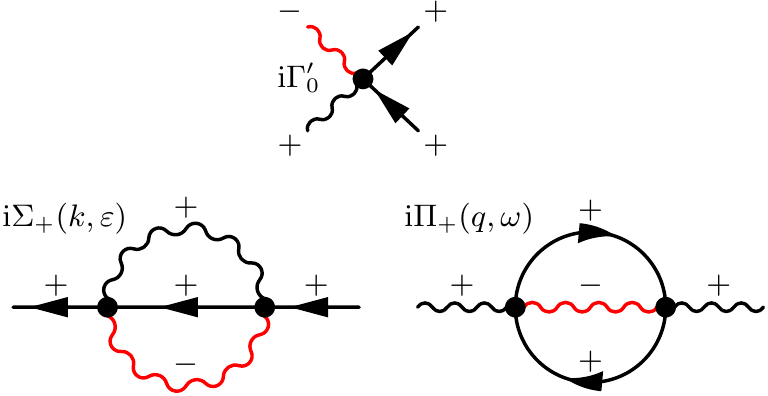}
\caption{(Color online) {Upper panel:} vertex corresponding to fermion-phonon interactions,
Eq.~(\ref{eq:lint}).
{Lower panel left:} diagram for the self-energy Keldysh matrix of right fermions,
$\Sigma_+(k,\varepsilon)$. {Lower panel right:} diagram for  the
self-energy Keldysh matrix of right phonons, $\Pi_+(q,\omega)$ (see the Supplemental Material). The solid line represents the fermionic propagator defined through Eq. \eqref{eq:lagrimp2nd}, and the wavy lines represent the right ($+$, black) or left ($-$, red) bosonic propagator defined through Eq. \eqref{eq:lchiral}.
\label{fig:FermionBosonSelf-energies-Gamma}}
\end{figure}

In the lowest (second) order, the interactions, corresponding to Eq.~(\ref{eq:lint})
modify fermionic and phononic propagators. The corresponding
self-energies $\Sigma_+(k,\varepsilon)$ and  $\Pi_+ (q,\omega$) 
are shown  in Fig.~\ref{fig:FermionBosonSelf-energies-Gamma}. Their imaginary
parts determine the rate of dissipative
processes. As expected, these processes consist of collisions of 
right-moving fermions with left-moving thermal phonons during which 
the phonons change their direction.

For short time scales, phonons do not participate in the dynamics and only
provide a damping mechanism for fermionic quasiparticles.  The dynamics at these time scales is described by a kinetic equation \cite{lifshitz1981physical} for small deviations
$\delta f_+(x,p;t)= f_+(x,p;t) - \bar{f}_+(p)$ of the fermionic distribution
function from its equilibrium form $\bar{f}_+(p)=1-2n_\mathrm{F} \big(E_+
(p)\big)\simeq \tanh(cp/2T) $. Making the scattering time approximation for
the linearized collision integral we have
\begin{eqnarray}
  \label{eq:kin_eq_F}
 \Big[ \dert  +\left(c+\frac{p}{m^*}\right)\derx\Big]\delta  f_+ = -\frac{\delta f_+}{\tau(p)}\, .
\end{eqnarray}
The time scale $\tau(p)$ is given by the imaginary part of the retarded component of the fermionic
self-energy matrix in the Keldysh space \footnote{ $\Sigma_+^\mathrm{R}$ is the upper-left element of the self-energy matrix $\Sigma_+$ in the Keldysh space. $\Pi_+^\mathrm{R}$ is the lower-left element of the bosonic polarization matrix $\Pi_+$ in the Keldysh space. For definitions and more details, see the Supplemental Material.}
\begin{eqnarray}
  \label{eq:tau_F_calc}
  \frac{1}{\tau(p)} = -2\im \Sigma_+^\mathrm{R} \big(p,E_+(p)\big) =
  \frac{\left[\Gamma'_0\right]^2 T^7}{2\pi {m^*}^2 c^6} 
  I\left(\frac{cp}{2T}\right)\,  
\end{eqnarray}
obtained from the corresponding diagram in Fig.~\ref{fig:FermionBosonSelf-energies-Gamma}. The dimensionless function
$I(y) =\int_{-\infty}^{+\infty}\!\! 
\dx\,x (x+2y)^4\big[\coth(x)-\tanh(x+y)\big]$  
has limiting behavior $I(y)=(\pi^6/64)(1 + (412/15 \pi^2) y^2) $ 
for  $y<1$. Consequently, 
the fermionic life-time is almost independent of momentum for $p<T/c$ 
and we obtain the result stated  in Eq.(\ref{eq:tauF}) for  $\tau=\tau(0)$. 

The DSF can be obtained \cite{lifshitz1981physical} by
solving Eq.~(\ref{eq:kin_eq_F}) with initial condition 
\begin{eqnarray}
  \label{eq:deltaf_init}
  \delta f_+ (x,p;0) = \frac{KT}{c}\frac{\der \bar{f}}{\der p}\delta(x)
  =\frac{K}{2\cosh^2\left(\frac{cp}{2T}\right)}\delta(x)\, ,
\end{eqnarray}
which yields 
\begin{eqnarray}
  \label{eq:sqw1}
  S(q,\omega) = \!\int\!\frac{\mathrm{d}p}{4\pi} 
  \frac{K\tau \cosh^{-2}(cp/2T)}{1+\tau^{2}\left(pq/m^*-(\omega-cq)\right)^2}\, .  
\end{eqnarray}
The integral in Eq.~(\ref{eq:sqw1}) 
contains the product of a Lorentzian with width 
$m^*/\tau q$ and the inverse square of a hyperbolic cosine with 
width $T/c$. 
For $q>q_c=m^* c/T\tau$, the Lorentzian is narrow and can be treated as a delta
function centered at $p=(m^*/q)(\omega-cq)$. 
The result of the integration is  
\begin{eqnarray}
\label{eq:sqwFlargeq}
  S(q,\omega) = \frac{K m^*}{4q}\cosh^{-2}
  \left(\frac{m^*c}{2Tq}(\omega-cq)\right)\, .
\end{eqnarray}
This expression justifies the width (\ref{eq:dwTG})
(for $m^*=m$, $K=1$) of  the DSF arising from 
the  thermal uncertainty of the initial wave  packet, 
Eq.~(\ref{eq:deltaf_init}).  

In the opposite
limit, $q<q_c$ the Lorentzian is a smooth function. Evaluating it at $p=0$ and
performing the remaining integration yields
\begin{eqnarray}
  \label{eq:sqwFsmallq}
  S(q,\omega) =\frac{1}{\pi}\frac{KT}{ c}
  \frac{\tau}{1+\tau^{2}(\omega-cq)^2}\, .   
\end{eqnarray}
In this regime, the width $\sim 1/\tau$ 
of the DSF is determined by collisions with thermal
phonons.

For long times, $t>\tau$, the fermionic quasiparticles decay, $\delta f_+
\to 0$, and the dynamics of phonons is governed by Eq.~(\ref{eq:kpz}).
The sole role of the fermionic quasiparticles is now  to provide a thermal 
bath for the phonons.  By calculating the diagram on the right in  
Fig.~\ref{fig:FermionBosonSelf-energies-Gamma},  
we get the dissipation and noise terms 
in Eq.~(\ref{eq:kpz})  from the imaginary part of the retarded phononic
self-energy $\Pi_+^\mathrm{R}(q,\omega)$ ~\cite{Note7}. The dissipation is proportional to the
diffusivity constant  
\begin{eqnarray}
  \label{eq:Diff}
  D = -\frac{\pi}{q^3}\im \Pi_+^\mathrm{R} (q,cq) = \frac{224}{15\pi^2}
  \left(\frac{c}{T}\right)^2 \frac{1}{\tau}\,  
\end{eqnarray}
confirming our statement that $\tau$ is the shortest time scale for the applicability of the
hydrodynamic approach.
The distribution of noise in Eq.~(\ref{eq:kpz})  
follows from the fluctuation-dissipation theorem. 

In conclusion, we have shown that the hydrodynamic description for 1D bosons
is only valid for times longer than $\tau$. The latter diverges as one
approaches the integrable point, and a non-hydrodynamic behavior due to
fermionic quasiparticles prevails. Using 
$\alpha = 12\ln(4/3)\hbar a^2\omega_\perp $ 
\cite{Mazets2008Breakdown,*PustilnikTan2010}, where $a$ is the scattering length and 
$\omega_\perp$ is the frequency of transverse confinement used to create a
one-dimensional system, we can recast Eq.~(\ref{eq:tauF}) into dimensionless 
form (reintroducing the Planck and Boltzmann constants),
\begin{eqnarray}
  \label{eq:tauF_dimless}
  \frac{\hbar}{\tau mc^2}=\frac{A}{K^3}\left(\frac{mc^2}{\hbar\omega_{\bot}}\right)^{2}\left(\frac{k_{B}T}{mc^2}\right)^{7}\, ,
\end{eqnarray}
where the numerical factor $A=9{\pi}^7 \left( \ln(4/3) \right)^2/2^{15}\simeq0.07$.
For experiments in Ref. ~\cite{Clement2011}, $K\sim10$, $mc^2/\hbar\sim5.5$kHz and $\omega_{\bot}\sim66$kHz, which results in $\tau$ of the order of tens of seconds even for $k_B T/{mc^2}=1$. Because of the $T^7$ dependence, decreasing the temperature will result in even longer relaxation times. This makes the fermionic ballistic result Eq.~(\ref{eq:sqwFlargeq}) the only one likely to be observed \footnote{The  collisional result,
Eq.~(\ref{eq:sqwFsmallq}), for these values of experimental parameters 
would require observation times similar to $\tau$.}.

In contrast to the purely hydrodynamic approach of Ref.~\cite{Andreev1980} and the
purely fermionic approach of Ref.~\cite{Rozhkov2005,*Rozhkov2008}, our
treatment is based on considering both phononic and fermionic excitations of
the liquid. This is justified \emph{a posteriori} as fermionic and phononic
fields fluctuate on very different time scales.

For the important special case of a linear spectrum, $1/m^*=0$, the scenario described
above is not correct as $1/\tau=0$ in this case. Nevertheless, the linearized (free) hydrodynamic description 
holds on all time scales and is at the origin of standard bosonization 
approach \cite{Gogolin2004Bosonization}. However, in this
case, the fermionic description is valid as well. The equivalence between the
two is due to the simple fact that fermionic wave packets do not disperse, and
their form can be parametrized by bosonic fields at all times.


This work originated from stimulating discussions with A. Lamacraft whom 
we wish to thank. We are also 
grateful to A. Kamenev, I. V. Lerner, J. M. F. Gunn, and L. I. Glazman
for fruitful comments and discussions. We also thank D.~Cl\'ement for his
update on  experimental advances.
M. A. and F. B. acknowledge the support of the University of Birmingham and EPSRC.

\bibliography{NLLrefDBshort}

\onecolumngrid
\newpage


\section*{Supplementary material}

# SUPPLEMENTAL MATERIAL

# Kinetics of mobile impurities and correlation functions in one-dimensional superfluids at finite temperature

M. Arzamasovs,[1] F. Bovo,[1] and D.M. Gangardt[1]

[1]*School of Physics and Astronomy, University of Birmingham, Edgbaston, Birmingham, B15 2TT, UK*

## I. DERIVATION OF FERMION-PHONON COUPLING

In a quantum liquid, a single impurity interacting with a phononic bath was studied in Ref. [1]. It was shown that it can be described by the universal Lagrangian

$$L = P\dot{X} - H(P,\Lambda) + \mathbf{\Lambda}^\dagger \frac{\mathrm{d}}{\mathrm{d}t}\boldsymbol{\chi}(X,t) \ . \tag{I.1}$$

The Hamiltonian $H(P,\Lambda)$ describes the dynamics of the impurity with momentum $P$ together with the depletion of the quantum liquid in its closest vicinity. In addition to the momentum $P$ and coordinate $X$, the state of such a system is described by two parameters $\mathbf{\Lambda}^\dagger = (\Lambda_+, \Lambda_-)$. Rewriting $2\Lambda_\pm = N/\sqrt{K} \pm (\Phi/\pi)\sqrt{K}$ one can associate the variable $N$ with the number of particles expelled from the vicinity of the impurity and $\Phi$ with the superfluid phase drop. Here the dynamics of the phonon field $\boldsymbol{\chi}(x,t)$ is governed by the Lagrangian $\mathcal{L}_{\mathrm{ph}} + \mathcal{L}'_{\mathrm{ph}}$, Eqs. (8) and (9) of main text. This picture applies equally to a foreign particle, for example an atom in a different hyperfine state considered in Refs.[1, 2] as well as to localized excitations in the superfluid itself, like grey solitons in BEC [3, 4]. The last term in Eq. (I.1) represents universal interactions of the impurity with the phononic fields $\boldsymbol{\chi}^\dagger = (\chi_+, \chi_-)$.

In the absence of phonons the depletion cloud variables take their equilibrium values $\mathbf{\Lambda} = \mathbf{\Lambda}^*(P)$ obtained from the conditions $\nabla_{\mathbf{\Lambda}} H = 0$. Equivalently

$$P = n\Phi^* - mN^*V(P) \ ,$$

$$\frac{\partial E(P)}{\partial n} = \frac{mc^2}{n}N^* - V(P)\Phi^* \ ,$$

which must be solved at constant $P$. Here $n$ is linear density of the background, $V(P) = \partial E/\partial P$ is the velocity and $E(P) = H(P, \mathbf{\Lambda}^*(P)) = cP + P^2/2m^*$ is the dispersion law of the impurity.

In the presence of long wavelength phonons the depletion cloud parameters change. We can, however, assuming that the change due to phononic background is small, expand $H(P,\Lambda)$ around $\Lambda^*$,

$$H(P, \mathbf{\Lambda}^* + \boldsymbol{\delta\Lambda}) \simeq E(P) + \boldsymbol{\delta\Lambda}^\dagger (2\pi \Gamma_P^{-1}) \boldsymbol{\delta\Lambda} \ , \tag{I.2}$$

and integrate out the fluctuations $\boldsymbol{\delta\Lambda}$. This results in effective Lagrangian of an impurity coupled to the bath of phonons:

$$L = P\dot{X} - E(P) + \mathbf{\Lambda}^{*\dagger}\frac{\mathrm{d}}{\mathrm{d}t}\boldsymbol{\chi}(X,t) + \frac{1}{8\pi}\left(\frac{\mathrm{d}}{\mathrm{d}t}\boldsymbol{\chi}^\dagger(X,t)\right)\Gamma_P\left(\frac{\mathrm{d}}{\mathrm{d}t}\boldsymbol{\chi}(X,t)\right) \ . \tag{I.3}$$

The matrix $\Gamma_P^{-1}$ is the Hessian,

$$\Gamma_P^{-1} = \begin{pmatrix} \Gamma_{P,++} & \Gamma_{P,+-} \\ \Gamma_{P,-+} & \Gamma_{P,--} \end{pmatrix} = \frac{1}{4\pi}\begin{pmatrix} H_{\Lambda_+\Lambda_+} & H_{\Lambda_+\Lambda_-} \\ H_{\Lambda_-\Lambda_+} & H_{\Lambda_-\Lambda_-} \end{pmatrix}_{\mathbf{\Lambda}=\mathbf{\Lambda}^*} \ ,$$

evaluated at constant $P$ (see also Ref. [1]). It characterizes the scattering rate of phonons by a moving impurity. It will be shown below that at small momenta $\mathbf{\Lambda}^*(P)$ is independent of $P$, so that the third term on the right of Eq. (I.3) is a total derivative and can be omitted. The full time derivative of phonon fields can be evaluated with the help of the equation of motion of the impurity, $\dot{X} = V(P)$, leading to $\mathrm{d}\boldsymbol{\chi}(X,t)/\mathrm{d}t \simeq (\partial_t + V(P)\partial_x)\boldsymbol{\chi}$. Then the Lagrangian becomes

$$L = P\dot{X} - E(P) + \frac{1}{8\pi}\left(\partial_t + V(P)\partial_x\right)\boldsymbol{\chi}^\dagger \Gamma_P \left(\partial_t + V(P)\partial_x\right)\boldsymbol{\chi} \ , \tag{I.4}$$

Evaluating the last term of Eq. (I.4) with the help of the equations of motion of phonons, $\partial_t \chi_\pm = \mp c \partial_x \chi_\pm$, we find

$$\frac{d}{dt}\chi_+ \Gamma_{P,++} \frac{d}{dt}\chi_+ = \left(\frac{P}{m^*}\right)^2 \partial_x \chi_+ \Gamma_{P,++} \partial_x \chi_+ \,,$$

$$\frac{d}{dt}\chi_- \Gamma_{P,-+} \frac{d}{dt}\chi_+ = \frac{d}{dt}\chi_+ \Gamma_{P,+-} \frac{d}{dt}\chi_- = \left(\frac{P}{m^*}\right)\left(2c + \frac{P}{m^*}\right) \partial_x \chi_+ \Gamma_{P,+-} \partial_x \chi_- \simeq \frac{2c}{m^*} P \partial_x \chi_+ \Gamma_{P,+-} \partial_x \chi_- \,, \quad (I.5)$$

$$\frac{d}{dt}\chi_- \Gamma_{P,--} \frac{d}{dt}\chi_- = \left(2c + \frac{P}{m^*}\right)^2 \partial_x \chi_- \Gamma_{P,--} \partial_x \chi_- \simeq 4c^2 \partial_x \chi_- \Gamma_{P,--} \partial_x \chi_- \,,$$

where we have simplified the expression leaving only the leading terms in $P/m^* c \ll 1$. The last term, being the largest, does not contribute, however to the physical quantities of interest (like the dissipation) since the corresponding scattering process is suppressed by the conservation of energy and momentum. We also neglect the first term as being small in the low energy limit. Moreover, since only vertices proportional to $\Gamma_{P,+-}$ and $\Gamma_{P,-+}$ are left, we denote $\Gamma_{P,+-} = \Gamma_{P,-+} \equiv \Gamma_P$. As it is shown in Sec. VI, for small momenta $\Gamma_P \simeq \Gamma'_0 P$.

So far only right-moving impurities have been considered. We can easily extend the discussion to left-moving impurities by means of the substitutions $E(P) \longrightarrow E_\pm(P) = \pm cP + P^2/2m^*$ and $V(P) \longrightarrow V_\pm(P) = \partial E_\pm/\partial P$ in the previous results, where $+$ and $-$ label right- and left-moving impurities. It can also be shown that $\Gamma^\pm_{P,++} = \Gamma^\mp_{P,--}$ and $\Gamma^\pm_{P,+-} = \Gamma^\mp_{P,+-}$. It follows that in second quantization the Lagrangian of right/left chiral fermions, Eq. (I.4), corresponds to the Lagrangian density $\mathcal{L}^\pm_F + \mathcal{L}^\pm_{\text{ph-F}}$ with

$$\mathcal{L}^\pm_F = \bar{\psi}_\pm \left[i\partial_t - \varepsilon_\pm(-i\partial_x)\right] \psi_\pm \,,$$
$$\mathcal{L}^\pm_{\text{ph-F}} = \pm \frac{1}{2\pi} \frac{c}{m^*} \Gamma'_0 \partial_x \chi_+^\alpha \partial_x \chi_-^\beta \left(\bar{\psi}_a (\overleftrightarrow{P})^2 \psi_b\right) \,. \quad (I.6)$$

Here $\varepsilon_+(p)$ is defined in Eq. (15) of the main text and we defined $\overleftrightarrow{P}$ as the symmetric version of the momentum operator, $\bar{\psi} \overleftrightarrow{P} \psi = \frac{i}{2} \left[(\partial_x \bar{\psi})\psi - \bar{\psi}(\partial_x \psi)\right]$, in order for it to be Hermitian. The respective term in Eq. (I.6) is explicitly given by $\bar{\psi}(\overleftrightarrow{P})^2 \psi = -\frac{1}{4}\left[(\partial_x^2 \bar{\psi})\psi - 2(\partial_x \bar{\psi})(\partial_x \psi) + \bar{\psi}(\partial_x^2 \psi)\right]$.

## II. RULES FOR FEYNMAN DIAGRAMS

In this section we derive the Feynman rules in the Keldysh formalism for the action

$$S = \int_\mathcal{C} dt \int dx \left(\mathcal{L}_F^+ + \mathcal{L}_F^- + \mathcal{L}_{\text{ph}} + \mathcal{L}'_{\text{ph}} + \mathcal{L}^+_{\text{ph-F}} + \mathcal{L}^-_{\text{ph-F}}\right) \,, \quad (II.1)$$

where $\mathcal{C}$ is the Keldysh contour. We perform Keldysh rotation

$$\begin{pmatrix} \chi^+ \\ \chi^- \end{pmatrix} = \begin{pmatrix} 1 & 1 \\ 1 & -1 \end{pmatrix} \begin{pmatrix} \chi^{\text{cl}} \\ \chi^{\text{q}} \end{pmatrix} \quad (II.2)$$

for bosonic fields and

$$\begin{pmatrix} \psi^+ \\ \psi^- \end{pmatrix} = \frac{1}{\sqrt{2}} \begin{pmatrix} 1 & 1 \\ 1 & -1 \end{pmatrix} \begin{pmatrix} \psi_1 \\ \psi_2 \end{pmatrix} \,, \qquad (\bar{\psi}^+, \bar{\psi}^-) = (\bar{\psi}_1, \bar{\psi}_2) \times \frac{1}{\sqrt{2}} \begin{pmatrix} 1 & -1 \\ 1 & 1 \end{pmatrix} \quad (II.3)$$

for fermionic ones. In these expressions $\pm$ superscripts denote the forward/backward parts of the Keldysh contour and we have suppressed the chirality indices as the expressions are identical for right/left moving fermions and bosons. As interactions between left and right fermions do not appear in the lowest order of perturbation theory, we concentrate here on right-moving fermions only and often omit the chirality index for clarity. In contrast, as both right and left moving phonons are excited via interaction with right-moving fermions, we indicate their chirality indices explicitly. We will also use the shorthand notation $k$ for $(k, \varepsilon)$ and $q$ for $(q, \omega)$ and omit the cut-off on both momenta. The propagators of fermions become matrices in Keldysh space

$$\xrightarrow{\pm}_{k} \quad = \mathcal{G}_\pm(k) = \begin{pmatrix} G^R_\pm(k,\varepsilon) & G^K_\pm(k,\varepsilon) \\ 0 & G^A_\pm(k,\varepsilon) \end{pmatrix} \,. \quad (II.4)$$

For phonons we have similarly

$$\underset{q}{\overset{+}{\sim\!\sim\!\sim\!\sim}} = \mathcal{D}_+(q) = \begin{pmatrix} D_+^K(q,\omega) & D_+^R(q,\omega) \\ D_+^A(q,\omega) & 0 \end{pmatrix} \tag{II.5}$$

and

$$\underset{q}{\overset{-}{\sim\!\sim\!\sim\!\sim}} = \mathcal{D}_-(q) = \begin{pmatrix} D_-^K(q,\omega) & D_-^R(q,\omega) \\ D_-^A(q,\omega) & 0 \end{pmatrix} . \tag{II.6}$$

The retarded and advanced propagators are given by

$$G_\pm^R(k,\varepsilon) = \left[G_\pm^A(k,\varepsilon)\right]^\dagger = \frac{1}{\varepsilon - \varepsilon_\pm(k) + i0} ,$$
$$D_\pm^R(q,\omega) = \left[D_\pm^A(q,\omega)\right]^\dagger = \frac{\pi}{\pm q(\omega \mp cq + i0)} . \tag{II.7}$$

In equilibrium the Keldysh propagators are given by

$$G_\pm^K(k,\varepsilon) = F_{f\pm}(k)\Delta_{f\pm}(k,\varepsilon) ,$$
$$D_\pm^K(q,\omega) = F_{b\pm}(q)\Delta_{b\pm}(q,\omega) , \tag{II.8}$$

where the density functions are $F_{f\pm}(k) = \tanh\left(\frac{\varepsilon_\pm(k)}{2T}\right)$ and $F_{b\pm}(q) = \coth\left(\frac{\omega_\pm(q)}{2T}\right)$ and we have defined

$$\Delta_{f\pm}(k,\varepsilon) = G_\pm^R(k,\varepsilon) - G_\pm^A(k,\varepsilon) = -2\pi i \delta\left(\varepsilon - \varepsilon_\pm(k)\right) ,$$
$$\Delta_{b\pm}(q,\omega) = D_\pm^R(q,\omega) - D_\pm^A(q,\omega) = \mp\frac{2\pi^2}{q}i\delta\left(\omega - \omega_\pm(q)\right) . \tag{II.9}$$

The vertex for the interaction between right chiral fermions and phonons in Eq. (I.6) is given by

$$\mathcal{L}_{\text{ph-F}}^+ = \frac{1}{2\pi}\frac{c}{m^*}\Gamma_0'\hat{\gamma}_{\alpha\beta}^{ab}\partial_x\chi_+^\alpha\partial_x\chi_-^\beta\left(\bar{\psi}_a(\overleftrightarrow{P})^2\psi_b\right)$$

$$\rightarrow \quad \begin{array}{c} q',\alpha,- \\ \\ q,\beta,+ \end{array} \!\!\!\!\!\!\!\!\!\!\!\!\!\!\!\!\! \begin{array}{c} k',a,+ \\ \\ k,b,+ \end{array} = \Gamma_{\beta\gamma}^{ac}(k',k;q',q) =$$

$$= \frac{i}{8\pi}\frac{c\Gamma_0'}{m^*}\gamma_{\alpha\beta}^{ab}qq'(k+k')^2(2\pi)^2\delta^2(k+q-k'-q') .$$

Here we have introduced Keldysh tensors $\hat{\gamma}_\alpha^{ab}$, $\hat{\gamma}_{\alpha\beta}^{ab}$, $a,b = 1,2$, $\alpha,\beta = \text{cl},\text{q}$ with components

$$\hat{\gamma}_{\text{cl}} = \begin{pmatrix} 1 & 0 \\ 0 & 1 \end{pmatrix}, \quad \hat{\gamma}_{\text{q}} = \begin{pmatrix} 0 & 1 \\ 1 & 0 \end{pmatrix}, \quad \hat{\gamma}_{\text{clcl}} = \hat{\gamma}_{\text{qq}} = \begin{pmatrix} 1 & 0 \\ 0 & 1 \end{pmatrix}, \quad \hat{\gamma}_{\text{clq}} = \hat{\gamma}_{\text{qcl}} = \begin{pmatrix} 0 & 1 \\ 1 & 0 \end{pmatrix} . \tag{II.10}$$

### III. CALCULATION OF SELF-ENERGIES

In this section we will calculate the self-energy of fermions and phonons at the lowest order in $\Gamma_0'$. In the Keldysh formalism they are given respectively by

$$\Sigma_\pm(k) = \begin{pmatrix} \Sigma_\pm^R(k,\varepsilon) & \Sigma_\pm^K(k,\varepsilon) \\ 0 & \Sigma_\pm^A(k,\varepsilon) \end{pmatrix}$$

and

$$\Pi_\pm(q) = \begin{pmatrix} 0 & \Pi_\pm^A(q,\omega) \\ \Pi_\pm^R(q,\omega) & \Pi_\pm^K(q,\omega) \end{pmatrix}$$

and the corresponding diagrams are shown in Fig. III.1.

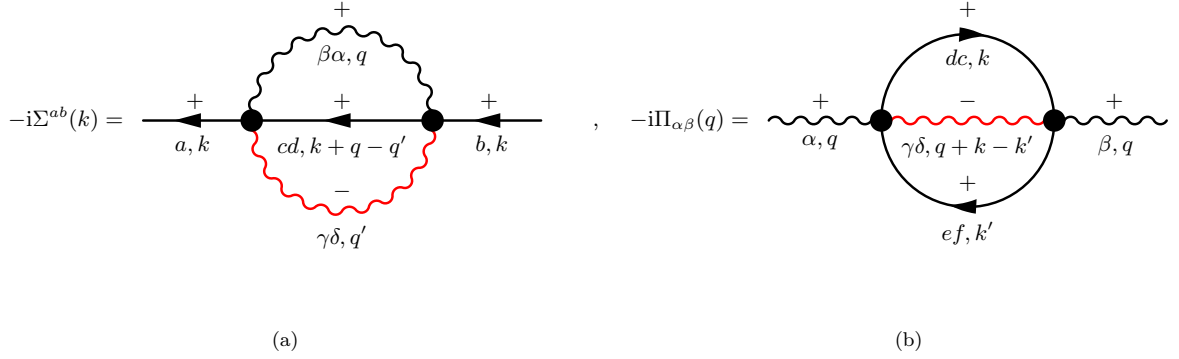

Figure III.1: Fermion and boson self-energies in the 2nd order perturbation theory.

### A. Fermion self-energy and life-time

The fermion self-energy is depicted in Fig. III.1(a) and reads

$$-i\Sigma^{ab}_{\pm}(k) = \int \frac{d^2q}{(2\pi)^2}\frac{d^2q'}{(2\pi)^2}\Gamma^{ac}_{\beta\gamma}(k+q-q',k;q',q)\, iG^{cd}_{\pm}(k+q-q')iD^{\gamma\delta}_{\mp}(q')$$
$$\times \Gamma^{db}_{\delta\alpha}(k,k+q-q';q,q')\, iD^{\alpha\beta}_{\pm}(q) \qquad\text{(III.1)}$$
$$= i\frac{1}{(8\pi)^2}\frac{\Gamma'^2_0 c^2}{(m^*)^2}\int \frac{d^2q}{(2\pi)^2}\frac{d^2q'}{(2\pi)^2}(qq')^2(2k+q-q')^4\left[\gamma^{ac}_{\beta\gamma}G^{cd}_{\pm}(k+q-q')D^{\gamma\delta}_{\mp}(q')\gamma^{db}_{\delta\alpha}D^{\alpha\beta}_{\pm}(q)\right],$$

where the factor 1/2 coming from the second order perturbation theory is cancelled by the factor 2 due to the symmetry by exchange of the vertices. The retarded component is given by

$$\Sigma^{R}_{\pm}(k) = \Sigma^{11}_{\pm}(k) = -\frac{1}{(8\pi)^2}\frac{\Gamma'^2_0 c^2}{(m^*)^2}\int \frac{d^2q}{(2\pi)^2}\frac{d^2q'}{(2\pi)^2}(qq')^2(2k+q-q')^4$$
$$\times \left[G^{R}_{\pm}(k+q-q')\left(D^{K}_{\mp}(q')D^{K}_{\pm}(q) - (D^{R}_{\mp}(q')-D^{A}_{\mp}(q'))(D^{R}_{\pm}(q)-D^{A}_{\pm}(q))\right) + \right.$$
$$\left. + G^{K}_{\pm}(k+q-q')\left(D^{K}_{\mp}(q')D^{A}_{\pm}(q) + D^{R}_{\mp}(q')D^{K}_{\pm}(q)\right)\right].$$

The Keldysh component is given by

$$\Sigma^{K}_{\pm}(k) = \Sigma^{12}_{\pm}(k) = -\frac{1}{(8\pi)^2}\frac{\Gamma'^2_0 c^2}{(m^*)^2}\int \frac{d^2q}{(2\pi)^2}\frac{d^2q'}{(2\pi)^2}(qq')^2(2k+q-q')^4$$
$$\times \left[G^{K}_{\pm}(k+q-q')\left(D^{K}_{\mp}(q')D^{K}_{\pm}(q) - (D^{R}_{\mp}(q')-D^{A}_{\mp}(q'))(D^{R}_{\pm}(q)-D^{A}_{\pm}(q))\right) + \right.$$
$$- D^{K}_{\mp}(q')\left(G^{R}_{\pm}(k+q-q') - G^{A}_{\pm}(k+q-q')\right)\left(D^{R}_{\pm}(q)-D^{A}_{\pm}(q)\right) +$$
$$\left. + D^{K}_{\pm}(q)\left(G^{R}_{\pm}(k+q-q') - G^{A}_{\pm}(k+q-q')\right)\left(D^{R}_{\mp}(q')-D^{A}_{\mp}(q')\right)\right].$$

In order to calculate the life-time and the kinetic equation we need the quantities

$$\Sigma^{K}_{\pm}(k) = O_f[\mathcal{K}_{f\pm}](k),\quad \mathcal{K}_{f\pm}(k,k';q,q') = F_{f\pm}(k')\left(F_{b\mp}(q')F_{b\pm}(q)-1\right) - F_{b\mp}(q') + F_{b\pm}(q),$$
$$2i\Im\Sigma^{R}_{\pm}(k) = O_f[\mathcal{T}_{f\pm}](k),\quad \mathcal{T}_{f\pm}(k,k';q,q') = F_{b\mp}(q')F_{b\pm}(q) - 1 + F_{f\pm}(k')\left(F_{b\pm}(q) - F_{b\mp}(q')\right),$$

where, employing Eq. (II.8), we defined

$$O_{f\pm}[\mathcal{F}](k) = -\frac{1}{(8\pi)^2}\frac{\Gamma'^2_0 c^2}{(m^*)^2}\int \frac{d^2q}{(2\pi)^2}\frac{d^2q'}{(2\pi)^2}d^2k'\,(qq')^2(k+k')^4\,\delta^2_{\pm}(k+q-q'-k')$$
$$\times \Delta_{f\pm}(k')\Delta_{b\mp}(q')\Delta_{b\pm}(q)\mathcal{F}_{f\pm}(k,k';q,q'). \qquad\text{(III.2)}$$

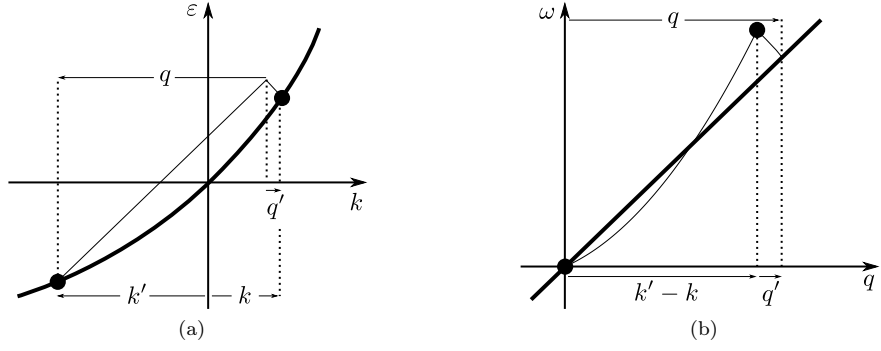

Figure III.2: Kinematics of the scattering processes described by the diagrams in Fig. (III.1). Given the weak non-linearity ($\sim 1/m^*$) of the fermionic spectrum, the momentum $q'$ of the left-moving phonon is parametrically small with respect to $q$.

Here we imposed the mass shell condition on the external energy as well since the quantities $O_f$'s that we will consider will always appear multiplying $\Delta$. The subscript $\pm$ of the delta function in $O_f$ refers to the energy conservation for right/left chiral fermions. Substituting the explicit values $\Delta_{f\pm}$, $\Delta_{b\pm}$, Eq. (II.9), and integrating over $d\omega$, $d\omega'$, $d\varepsilon'$ we find

$$O_{f\pm}[\mathcal{F}](k) = \frac{\mathrm{i}}{128\pi} \frac{{\Gamma'_0}^2 c^2}{(m^*)^2} \int dq\, dq'\, dk'\, \delta(\varepsilon_\pm(k) + \omega_\pm(q) - \omega_\mp(q') - \varepsilon_\pm(k'))\delta(k+q-q'-k')qq'(k+k')^4 \mathcal{F}_{f\pm}(k,k';q,q') \ .$$

By integrating over $dk'$, making the shift $q' = q - \tilde{q}$ and integrating $dq$ out, we find

$$O_{f\pm}[\mathcal{F}](k) = \pm\frac{\mathrm{i}}{128\pi} \frac{{\Gamma'_0}^2}{(2m^*)^3} \int dq\, q^2\, (2k+q)^5 \left[1 \pm \frac{2k+q}{4m^*c}\right] \mathcal{F}_{f\pm}\left(k, k+q; q\left[1 \pm \frac{2k+q}{4m^*c}\right], \pm q\frac{2k+q}{4m^*c}\right) \ ,$$

where in the last line we renamed the dummy variable $\tilde{q}$ to $q$. At the lowest order in $k, q \ll m^*c$ (i.e., non-linearity of the spectrum) we have

$$O_{f\pm}[\mathcal{F}](k) \simeq \pm\frac{\mathrm{i}}{128\pi} \frac{{\Gamma'_0}^2}{(2m^*)^3} \int dq\, q^2\, (2k+q)^5 \mathcal{F}_{f\pm}\left(k, k+q; q, \pm q\frac{2k+q}{4m^*c}\right) \ . \tag{III.3}$$

The last argument of $\mathcal{F}_{f\pm}$ is the momentum of the left-moving (right-moving) phonon in the scattering process which is small due to the conservation of energy and momentum. Indeed it would be zero in the case of a linear dispersion ($1/m^* = 0$), and the addition of a weak non-linearity ($q, k \ll m^*c$) makes it parametrically small (see Fig. III.2a).

The lifetime of fermions is given by

$$\frac{1}{\tau(k)} = -2\Im\Sigma^{\mathrm{R}}(k) = \mathrm{i}O_f[\mathcal{T}_{f\pm}](k) \ ,$$

where we omitted the chiral indices as the result will be independent of them. At equilibrium

$$\mathcal{T}_{f\pm}(k,k';q,q') = \left[\coth\left(\frac{cq'}{2T}\right) + \coth\left(\frac{cq}{2T}\right)\right]\left[\tanh\left(\frac{ck'}{2T}\right) - \coth\left(\frac{c(q+q')}{2T}\right)\right] \ .$$

In the limit $k, q \ll m^*c$, the terms with the divergent factor $\coth(cq'/2T)$ dominate as a consequence of the conservation of energy and momentum and, substituting into Eq. (III.3), the expression simplifies to

$$\frac{1}{\tau(k)} = \frac{1}{2\pi} \frac{T^7 {\Gamma'_0}^2}{m^{*2} c^6} I\left(\frac{ck}{2T}\right) = \frac{\pi^5}{128} \frac{T^7 {\Gamma'_0}^2}{(m^*c^2)^2 c^2} + \left(\frac{103\pi^3}{1920} \frac{T^5 {\Gamma'_0}^2}{(m^*c^2)^2}\right) k^2 + o\left(k^4\right) \ ,$$

$$I(y) = \int_{-\infty}^{+\infty} dx\, x\, (x+2y)^4 \left[\coth(x) - \tanh(x+y)\right] \ . \tag{III.4}$$

In the leading order in $k$ it results in Eq. (4) of the main text.

The life-time of excitations in the system of nonlinear interacting fermions was investigated in [5, 6]. Their result for a non-thermal particle scales as $k^8$ where $k$ is the momentum of the quasiparticle above Fermi surface. Substituting signum functions for distribution functions,

$$\tanh\left(\frac{E}{2T}\right) \to \text{sgn}(E),$$

$$\coth\left(\frac{\omega}{2T}\right) \to \text{sgn}(\omega),$$

we can verify this zero temperature result.

$$\frac{1}{\tau(k)} = -\frac{\Gamma_0'^2 c}{256\pi m^{*2}} \int dq dq' \delta\left(q' - \frac{q(2k+q)}{4m^*c}\right) \times$$

$$\times qq'(2k+q)^4 \left[\text{sgn}(-cq')\{\text{sgn}(cq) - \text{sgn}(ck+cq)\} + \{\text{sgn}(ck+cq)\text{sgn}(cq) - 1\}\right],$$

where we have neglected the curvature of the spectrum, as well as $q'$, in the distribution functions, but retained it inside the energy delta–function. We have also conveniently rearranged the distribution functions such that the expressions in curly brackets are non–zero only when $-k < q < 0$. Since $\text{sgn}(-cq') = \text{sgn}(-(2k+q)q) = 1$ in that region, the value of the expression in square brackets gives a factor of $-4$ and the decay rate is given by

$$\tau^{-1}(k) = \frac{\Gamma_0'^2 c^2}{128\pi m^{*2}} \int_{-k}^{0} dq \frac{q^2(2k+q)^5}{2m^*c^2} = \frac{73\Gamma_0'^2 k^8}{14336\pi m^{*3}}.$$

In the case of *thermal* quasiparticles the temperature dependence $T^7$, see Eq. (III.4), originates from the fact that the energy of the left moving phonon $cq'$ is controlled by *first* power of temperature $T$, in contrast to *second* power of the incoming momentum in Refs. [5, 6].

### B. Boson self-energy and diffusivity

The boson self-energy is depicted in Fig. III.1(b) and reads

$$-i\Pi_{\pm,\alpha\beta}(q) = -\int \frac{d^2k}{(2\pi)^2} \frac{d^2k'}{(2\pi)^2} \Gamma^{de}_{\alpha\gamma}(k', k; q+k-k', q) \, iG^{ef}_{\pm}(k') iD^{\gamma\delta}_{\mp}(q+k-k')$$

$$\times \Gamma^{fc}_{\delta\beta}(k, k'; q, q+k-k') \, iG^{cd}_{\pm}(k)$$

$$= -\frac{i}{(8\pi)^2} \frac{\Gamma_0'^2 c^2}{(m^*)^2} \int \frac{d^2k}{(2\pi)^2} \frac{d^2k'}{(2\pi)^2} q^2(q+k-k')^2 (k+k')^4 \left[\gamma^{de}_{\alpha\gamma} G^{ef}_{\pm}(k') D^{\gamma\delta}_{\mp}(q+k-k') \gamma^{fc}_{\delta\beta} \mathcal{G}_{\pm}{}^{cd}(k)\right],$$

$$\text{(III.5)}$$

where the minus sign in the first line is given by the fermionic loop and, again, the factor $1/2$ coming from the second order perturbation theory is cancelled by the factor $2$ due to the symmetry by exchange of the vertices. The retarded component is given by

$$\Pi^R_{\pm}(q) = \Pi_{\pm,\text{qcl}}(q) = \frac{1}{(8\pi)^2} \frac{\Gamma_0'^2 c^2}{(m^*)^2} \int \frac{d^2q}{(2\pi)^2} \frac{d^2q'}{(2\pi)^2} q^2(q+k-k')^2 (k+k')^4$$

$$\times \left[D^K_{\mp}(q+k-k')\left(G^R_{\pm}(k')G^K_{\pm}(k) + G^K_{\pm}(k')G^A_{\pm}(k)\right) + \right.$$

$$\left. + D^R_{\mp}(q+k-k')\left(G^K_{\pm}(k')G^K_{\pm}(k) - \left(G^R_{\pm}(k') - G^A_{\pm}(k')\right)\left(G^R_{\pm}(k) - G^A_{\pm}(k)\right)\right)\right].$$

The Keldysh component is given by

$$\Pi^K_{\pm}(q) = \Pi_{\pm,\text{qq}}(q) = \frac{1}{(8\pi)^2} \frac{\Gamma_0'^2 c^2}{(m^*)^2} \int \frac{d^2k}{(2\pi)^2} \frac{d^2k'}{(2\pi)^2} q^2(q+k-k')^2 (k+k')^4$$

$$\times \left[D^K_{\mp}(q+k-k')\left(G^K_{\pm}(k')G^K_{\pm}(k) - \left(G^R_{\pm}(k') - G^A_{\pm}(k')\right)\left(G^R_{\pm}(k) - G^A_{\pm}(k)\right)\right) + \right.$$

$$\left. - G^K_{\pm}(k')\left(\left(D^R_{\mp}(q+k-k') - D^A_{\mp}(q+k-k')\right)\left(G^R_{\pm}(k) - G^A_{\pm}(k)\right)\right) + \right.$$

$$\left. + G^K_{\pm}(k)\left(D^R_{\mp}(q+k-k') - D^A_{\pm}(q+k-k')\right)\left(G^A_{\pm}(k') - G^R_{\pm}(k')\right)\right],$$

where we used the causality structure of the Green's functions. In order to calculate the diffusion coefficient and the effective action of bosons we need the quantities

$$\Pi^K_\pm(q) = O_b[\mathcal{K}_{b\pm}](q), \quad \mathcal{K}_{b\pm}(k, k'; q, q') = F_\mp(q') (F_{f\pm}(k')F_{f\pm}(k) - 1) - F_{f\pm}(k') + F_{f\pm}(k),$$

$$2i\Im\Pi^R_\pm(q) = O_b[\mathcal{T}_{b\pm}](q), \quad \mathcal{T}_{f\pm}(k, k'; q, q') = F_\mp(q') (F_{f\pm}(k) - F_{f\pm}(k')) + F_{f\pm}(k')F_{f\pm}(k) - 1,$$

where we used Eq. (II.8) and defined

$$O_{b\pm}[\mathcal{F}](q) = \frac{1}{(8\pi)^2} \frac{\Gamma_0'^2 c^2}{(m^*)^2} \int \frac{d^2k}{(2\pi)^2} \frac{d^2k'}{(2\pi)^2} d^2q' q^2 (q')^2 (k + k')^4 \delta^2_\pm(q + k - k' - q')$$
$$\times \Delta_\mp(q')\Delta_{f\pm}(k')\Delta_{f\pm}(k)\mathcal{F}_{b\pm}(q, q'; k, k').$$

Substituting the explicit values of $\Delta_{f\pm}(k)$ and $\Delta_{b\pm}(q)$, Eq. (II.9), and integrating over $d\varepsilon$, $d\varepsilon'$, $d\omega'$ we find

$$O_{b\pm}[\mathcal{F}](q) = \mp\frac{i}{2(8\pi)^2} \frac{\Gamma_0'^2 c^2}{(m^*)^2} \int dk dk' dq' q^2 q' (k + k')^4 \delta(\omega_\pm(q) + \varepsilon_\pm(k) - \varepsilon_\pm(k') - \omega_\mp(q'))\delta(q + k - k' - q')\mathcal{F}_{b\pm}(q, q'; k, k').$$

Integrating over $dq'$, making the change of variables $k \to (k + k')/2$, $k' \to (k - k')/2$ and integrating over the new $dk'$, the integral reduces to

$$O_{b\pm}[\mathcal{F}](q) = -\frac{i}{(8\pi)^2} \frac{\Gamma_0'^2}{32(m^*)^3} q^3 \int dk k^5 \left|\frac{1}{1 \pm \frac{k}{4m^*c}}\right| \left(\frac{1}{1 \pm \frac{k}{4m^*c}}\right)$$
$$\times \mathcal{F}_{b\pm}\left(q, \pm\frac{qk}{4m^*c}\left(\frac{1}{1 \pm \frac{k}{4m^*c}}\right); \frac{1}{2}\left[k - q\left(\frac{1}{1 \pm \frac{k}{4m^*c}}\right)\right], \frac{1}{2}\left[k + q\left(\frac{1}{1 \pm \frac{k}{4m^*c}}\right)\right]\right).$$

At the lowest order in $k, q \ll m^*c$, we have

$$O_{b\pm}[\mathcal{F}](q) = -\frac{i}{256\pi^2} \frac{\Gamma_0'^2}{(2m^*)^3} q^3 \int dk k^5 \mathcal{F}_{b\pm}\left(q, \pm\frac{qk}{4m^*c}; \frac{k-q}{2}, \frac{k+q}{2}\right). \tag{III.6}$$

Here the second argument of $\mathcal{F}_{b\pm}$ represents the momentum of the left-moving (right-moving) boson in the scattering process. The momentum is parametrically small for the same reason explained in the fermionic case (see Fig. III.2b).

The diffusivity of phonons is given by

$$D(q) = -\left(\pm\frac{\pi}{q}\right)\frac{1}{q^2}\Im\Pi^R(q) = \pm i\frac{\pi}{q^3}O_b[\mathcal{T}_{b\pm}](k),$$

where we omitted the chiral indices as the result will be independent of them. At equilibrium

$$\mathcal{T}_{b\pm}(q, q'; k, k') = \left[\coth\left(\frac{cq'}{2T}\right) - \coth\left(\frac{c(k' - k)}{2T}\right)\right]\left[\tanh\left(\frac{ck'}{2T}\right) - \tanh\left(\frac{ck}{2T}\right)\right].$$

In the limit $k, q \ll m^*c$, the terms with the divergent factor $\coth\left(\frac{cq'}{2T}\right)$ dominate as a consequence of the conservation of energy and momentum and, substituting into Eq. (III.6), the expression simplifies to

$$D(q) = \frac{1}{2\pi} \frac{T\Gamma_0'^2}{256(m^*)^2} \frac{1}{q} \int dk k^4 \left[\tanh\left(\frac{c}{2T}\frac{k+q}{2}\right) - \tanh\left(\frac{c}{2T}\frac{k-q}{2}\right)\right].$$

Using Eq. (A.3) for the second integral, we find [11]

$$D(q) = \frac{7\pi^3}{120} \frac{T^5 \Gamma_0'^2}{(m^*c^2)^2} + o(q^2). \tag{III.7}$$

In the leading order in $q$ it results in Eq. (24) of the main text.

## IV. PHONON EFFECTIVE ACTION

Including the self energies, the Lagrangian of right phonons reads

$$L[\chi] = \frac{1}{4} \int \frac{d^2q}{(2\pi)^2} \left\{ 2\chi_+^{q\dagger} \left( \left[D_+^{-1}\right]^R - i\Im\Pi^R \right) \chi_+^{cl} - \chi_+^{q\dagger} \Pi^K \chi_+^q \right\} .$$

By means of the FDT we can obtain the Keldysh component of the self-energy,

$$\Pi^K(q) = \coth\left(\frac{cq}{2T}\right) \left(\Pi^R(q) - \Pi^A(q)\right) = \coth\left(\frac{cq}{2T}\right) \left[-i\left(\frac{q}{\pi}\right)\tau^{-1}(q)\right] = -\frac{2}{\pi} iDq^3 \coth\left(\frac{cq}{2T}\right) ,$$

where $D$ is the constant part of $D(q)$. The Lagrangian becomes

$$L[\chi] = \frac{1}{2\pi} \int \frac{d^2q}{(2\pi)^2} \left\{ \chi_+^{q\dagger} \left(q(\omega - cq) + iDq^3\right) \chi_+^{cl} + \chi_+^{q\dagger} \left(iDq^3 \coth\left(\frac{cq}{2T}\right)\right) \chi_+^q \right\} .$$

The Fourier transform gives [7]

$$L[\chi] = \frac{1}{2\pi} \int dx \left\{ -\partial_x \chi_+^q \left((\partial_t + c\partial_x) - D\partial_x^2\right) \chi_+^{cl} + \right.$$
$$\left. + i\frac{D}{2} \left[ \frac{2T}{c} (\partial_x \chi_+^q)^2 + \frac{\pi T^2}{2} \int dx' \frac{(\partial_x \chi_+^q(x) - \partial_{x'} \chi_+^q(x'))^2}{\sinh^2\left[\pi \frac{T}{C}(x - x')\right]} \right] \right\} .$$

In the semiclassical limit the last term is zero and the remaining quadratic term in the quantum field can be split by means of the Hubbard-Stratonovich transformation and the Lagrangian becomes

$$L[\chi] = -\frac{1}{2\pi} \int dx \left\{ \partial_x \chi_+^q \left((\partial_t + c\partial_x) - D\partial_x^2\right) \chi_+^{cl} - \partial_x \chi_+^q \xi \right\} ,$$

where $\xi$ is a Gaussian random force with second momentum given by $\langle \xi(x)\xi(x') \rangle = 4\pi \frac{DT}{c} \delta(x - x')$. Finally, reintroducing the chiral indices, the equation of motion is

$$(\partial_t \pm c\partial_x) \chi_\pm^{cl} = \frac{(\partial_x \chi_\pm^{cl})^2}{2m^*} + D\partial_x^2 \chi_\pm^{cl} + \xi ,$$

where we introduced the non-linear term and obtain Eq. (11) of the main text.

## V. KINETIC EQUATIONS FOR FERMIONS

The kinetic equations for fermions in the slightly inhomogeneous case is given by [7]

$$[\partial_t + (\pm c + k/m) \partial_x] F_{f\pm}(x, k) = iI_{f\pm}(x, k) ,$$

where $I$ is the collisional integral, given by

$$I_{f\pm}(x, k) = \Sigma_\pm^K(x, k) - 2iF_{f\pm}(x, k)\Im\Sigma_\pm^R(x, k)$$

and the functional dependence on $F_{f\pm}$ and $F_{b\pm}$ has been omitted. Since the system is slightly inhomogeneous, we can neglect the spatial derivatives in the collision integrals, and use the results of Sec. III. In terms of Eq. (III.2), $I_{f\pm} = O_{f\pm}[\mathcal{I}_{f\pm}]$, with

$$\mathcal{I}_{f\pm} = \mathcal{K}_{f\pm} - F_{f\pm}\mathcal{T}_{f\pm} .$$

To study the system in a state out of equilibrium we substitute $F_{f\pm}(x, k) = \tanh\left(\frac{\varepsilon_\pm(k)}{2T}\right) + f_{f\pm}(x, k)$, where $f_{f\pm}(x, k)$ is the displacement from the equilibrium distribution, and consider only the linear approximation in the variations. Keeping only the linear terms we find the linearized kinetic equation

$$(\partial_t + v_\pm(k)\partial_x) f_{f\pm}(x, k) = -\frac{f_{f\pm}(x, k)}{\tau(k)} + \int dq G(k, q) f_{f\pm}(x, k + q) , \quad \text{(V.1)}$$

where

$$G(k,q) = \frac{1}{128\pi} \frac{T\Gamma_0'^2}{m^{*2}} q(2k+q)^4 \left( \coth\left(\frac{cq}{2T}\right) + \tanh\left(\frac{ck}{2T}\right) \right).$$

The kinetic equation (18) in the main text is considered in the scattering time approximation, that is, only the first term in the right hand side of Eq. (V.1) is kept, with the term containing $G(k,q)$ being negligible.

## VI. CALCULATION OF THE BACK-SCATTERING AMPLITUDE

The backscattering rate can be calculated as the power series expansion in $P$, up to the first order for low momenta

$$\Gamma_P = \Gamma_{+-} = \Gamma_{-+} = -\frac{1}{c} \left( \Phi \frac{\partial N}{\partial P} - N \frac{\partial \Phi}{\partial P} + \frac{\partial N}{\partial n} \right) \approx \Gamma_0 + \Gamma_0' P.$$

Expanding $N(P)$ and $\Phi(P)$ in Taylor series

$$N(P) = N_0 + N_1 P + N_2 P^2 + \ldots,$$

$$\Phi(P) = \Phi_0 + \Phi_1 P + \Phi_2 P^2 + \ldots,$$

and substituting in the expression for $\Gamma_P$ one obtains

$$\Gamma_0 = -\frac{1}{c} \left( \Phi_0 N_1 - N_0 \Phi_1 + \frac{\partial N_0}{\partial n} \right)$$

and

$$\Gamma_0' = -\frac{1}{c} \left( 2\Phi_0 N_2 - 2N_0 \Phi_2 + \frac{\partial N_1}{\partial n} \right).$$

The equilibrium values of $N^*$, $\Phi^*$ (we drop asterisks from now on) can be obtained by inverting the equations for $N$ and $\Phi$:

$$N(P) = \frac{1}{m} \frac{V(P)P + n\partial E(P)/\partial n}{c^2 - V^2(P)}$$

$$\Phi(P) = \frac{P}{n} + \frac{mV(P)}{n} N(P).$$

Using the quadratic dispersion, one has

$$V(P) = c + \frac{P}{m^*}$$

$$\frac{\partial E(P)}{\partial n} = \frac{\partial c}{\partial n} P - \frac{P^2}{2(m^*)^2} \frac{\partial m^*}{\partial n}.$$

Therefore,

$$N_0 = -\frac{m^*}{2m} \left( 1 + \frac{n}{c} \frac{\partial c}{\partial n} \right) = -\sqrt{K}$$

$$\Phi_0 = -\frac{m^* c}{2n} \left( 1 + \frac{n}{c} \frac{\partial c}{\partial n} \right) = -\frac{\pi}{\sqrt{K}}.$$

Here we have used the expression for effective mass $m^*$ in the Luttinger liquid cited in [8, 9],

$$\frac{1}{m^*} = \frac{c}{K}\frac{\partial}{\partial \mu}(c\sqrt{K}) = \frac{1}{2m\sqrt{\frac{\pi n}{mc}}}\left(1 + \frac{n}{c}\frac{\partial c}{\partial n}\right).$$

The first order coefficients are:

$$N_1 = \frac{1}{2mc}\left(\frac{n}{2m^*}\frac{\partial m^*}{\partial n} + \frac{n}{2c}\frac{\partial c}{\partial n} - \frac{1}{2}\right)$$

$$\Phi_1 = \frac{1}{2n}\left(\frac{n}{2m^*}\frac{\partial m^*}{\partial n} - \frac{n}{2c}\frac{\partial c}{\partial n} + \frac{1}{2}\right).$$

The expression for $\Gamma_0$ then simplifies to

$$\Gamma_0 = -\frac{m^*}{4mcn}\left(1 - \left(\frac{n}{c}\frac{\partial c}{\partial n}\right)^2\right) - \frac{1}{c}\frac{\partial N_0}{\partial n}.$$

Substituting the expression for $m^*$ and using

$$-\frac{\partial N_0}{\partial n} = \frac{\partial}{\partial n}\sqrt{\frac{\pi n}{mc}} = \sqrt{\frac{\pi}{mnc}}\left(1 - \frac{n}{c}\frac{\partial c}{\partial n}\right),$$

one obtains that $\Gamma_0$ vanishes identically,

$$\Gamma_0 = 0.$$

$\Gamma_0'$ takes the form

$$\Gamma_0' = -\frac{1}{c}\left(\left(1 + \frac{n}{c}\frac{\partial c}{\partial n}\right)\frac{N_1}{n} + \frac{\partial N_1}{\partial n}\right), \tag{VI.1}$$

where the second order coefficients cancel because of the relation between $N(P)$ and $\Phi(P)$:

$$\Phi_2 = \frac{mc}{n}N_2 + \frac{m}{m^*n}N_1.$$

For

$$\mu(n) = gn, \qquad c(n) = \sqrt{\frac{gn}{m}},$$

the following expression for $\Gamma_0'$ is obtained:

$$\Gamma_0' = \frac{1}{16mc^2n}. \tag{VI.2}$$

However, we expect $\Gamma_P$ to vanish identically for integrable systems, such as Lieb–Liniger model. The above non-zero result is an artefact of the mean-field approximation used for $\mu(n)$. On the other hand, the contribution of the three-body collisions, for which

$$\mu(n,\alpha) = gn + \frac{\alpha}{2}n^2, \qquad c(n,\alpha) = \sqrt{\frac{gn + \alpha n^2}{m}} \approx \sqrt{\frac{gn}{m}}\left(1 + \frac{\alpha n}{2g}\right),$$

is not expected to vanish. Substituting the expression for sound velocity into Eqs. (VI.1), and subtracting Eq. (VI.2), we finally obtain

$$\Gamma_0' \approx -\frac{1}{48}\frac{\alpha n}{m^2 c^4}$$

up to the first order in $\alpha$.

## Appendix A: Identities

*Hyperbolic tangent and cotangent identities*

$$1 - \tanh(a)\tanh(b) = \coth(a-b)\left[\tanh(a) - \tanh(b)\right] \tag{A.1}$$

$$1 - \coth(a)\coth(b) = \coth(a-b)\left[\coth(a) - \coth(b)\right] \tag{A.2}$$

*Integral of difference of hyperbolic tangents*

$$I(y) \equiv \int_{-\infty}^{\infty} dx\, x^{n-1} \left[\tanh(x+y) - \tanh(x-y)\right] =$$

$$= \frac{4}{n}\left[y^n + \sum_{m=1}^{(n-1)/2} \binom{n}{2m} \pi^{2m}\left(1 - 2^{1-2m}\right)|B_{2m}|\, y^{n-2m}\right],\quad n \text{ odd}$$

Here $B_n$ are the Bernoulli numbers.

*Proof.* Integrating by parts we obtain

$$I(y) = \frac{1}{n}\int_{-\infty}^{\infty} dx\, x^n \left[\frac{1}{\cosh^2(x-y)} - \frac{1}{\cosh^2(x+y)}\right]$$

where the boundary term is zero as

$$\lim_{\Lambda \to \infty}\left\{\frac{x^n}{n}\left[\tanh(x+y) - \tanh(x-y)\right]\right\}_{-\Lambda}^{\Lambda} = \lim_{\Lambda \to \infty}\left[4\frac{\sinh(2y)}{1 + 2\cosh(2y) + e^{-4\Lambda}}e^{-2\Lambda}\right]_{-\Lambda}^{\Lambda} = 0$$

By shifting the variable as $x \to x \pm y$ the integral reads

$$I(y) = \frac{1}{n}\int_{-\infty}^{\infty} dx\, [(x+y)^n - (x-y)^n]\frac{1}{\cosh^2(x)}$$

Since the integrand is even we can write twice the integral from zero to infinity. Expressing the binomials with the help of the binomial theorem we find

$$I(y) = \frac{2}{n}\sum_{k=0}^{n}\binom{n}{k}\left(y^{n-k} - (-y)^{n-k}\right)\int_0^{\infty} dx\,\frac{x^k}{\cosh^2(x)}$$

The term in parenthesis is zero for $k$ odd and equal to $2y^{n-k}$ for $k$ even. Therefore, putting $k = 2m$

$$I(y) = \frac{4}{n}\sum_{m=0}^{(n-1)/2}\binom{n}{2m}y^{n-2m}\int_0^{\infty} dx\,\frac{x^{2m}}{\cosh^2(x)}$$

Using formulas 3.511.8 and 3.527.5 of Ref. [10] we find the result. □

In the case in which a positive constant $A$ multiplies the argument of the hyperbolic tangent, the integral becomes

$$I(y) \equiv \int_{-\infty}^{\infty} dx\, x^{n-1}\left[\tanh(A(x+y)) - \tanh(A(x-y))\right] =$$

$$= \frac{4}{n}y^n\left[1 + \sum_{m=1}^{(n-1)/2}\binom{n}{2m}\left(\frac{\pi}{A}\right)^{2m}\left(1 - 2^{1-2m}\right)|B_{2m}|\, y^{-2m}\right],\quad n \text{ odd} \tag{A.3}$$



\end{document}